\documentclass[a4paper,twocolumn,accepted=2019-09-02]{quantumarticle}
\pdfoutput=1

\usepackage{mathtools}
\usepackage{amsthm,amsmath,amssymb,amsfonts}
\usepackage{bbm}
\usepackage{bm}
\usepackage{dsfont}
\usepackage[normalem]{ulem}

\usepackage[numbers,sort&compress]{natbib}

\def\bra#1{\mathinner{\langle{#1}|}}
\def\ket#1{\mathinner{|{#1}\rangle}}
\def\braket#1{\mathinner{\langle{#1}\rangle}}
\newcommand{\ketbra}[2]{\mathinner{|{#1}\rangle\!\langle{#2}|}}

{\catcode`\|=\active 
  \gdef\Braket#1{\left<\mathcode`\|"8000\let|\BraVert {#1}\right>}}
\def\BraVert{\egroup\,\mid@vertical\,\bgroup}

\DeclareMathOperator{\Tr}{Tr}

\DeclareMathOperator{\re}{Re}
\DeclareMathOperator{\im}{Im}
\newcommand{\id}{\mathbbm{1}}
\newcommand{\idmini}{\scalebox{.7}{$\id$}}

\usepackage[utf8]{inputenc}
\usepackage[T1]{fontenc}
\usepackage[colorlinks]{hyperref}

\begin{document}

\title{Anomalous Weak Values Without Post-Selection}

\author{Alastair A.\ Abbott}
\thanks{A.A.A.\ and R.S.\ contributed equally to this work.}
\affiliation{Univ.\ Grenoble Alpes, CNRS, Grenoble INP, Institut N\'eel, 38000 Grenoble, France}
\orcid{0000-0002-2759-633X}

\author{Ralph Silva}
\affiliation{D\'epartement de Physique Appliqu\'ee, Universit\'e de Gen\`eve, 1211 Gen\`eve, Switzerland}
\affiliation{Institute for Theoretical Physics, ETH Zurich, 8093 Zurich, Switzerland}
\orcid{0000-0002-4603-747X}

\author{Julian Wechs}
\affiliation{Univ.\ Grenoble Alpes, CNRS, Grenoble INP, Institut N\'eel, 38000 Grenoble, France}
\orcid{0000-0002-0395-6791}

\author{Nicolas Brunner}
\affiliation{D\'epartement de Physique Appliqu\'ee, Universit\'e de Gen\`eve, 1211 Gen\`eve, Switzerland}
\orcid{0000-0002-2104-9191}

\author{Cyril Branciard}
\affiliation{Univ.\ Grenoble Alpes, CNRS, Grenoble INP, Institut N\'eel, 38000 Grenoble, France}
\orcid{0000-0001-9460-825X}

\date{September 2, 2019}

\begin{abstract}
A weak measurement performed on a pre- and post-selected quantum system can result in an average value that lies outside of the observable's spectrum. 
This effect, usually referred to as an ``anomalous weak value'', is generally believed to be possible only when a non-trivial post-selection is performed, i.e., when only a particular subset of the data is considered. 
Here we show, however, that this is not the case in general: in scenarios in which several weak measurements are sequentially performed, an anomalous weak value can be obtained without post-selection, i.e., without discarding any data. 
We discuss several questions that this raises about the subtle relation between weak values and pointer positions for sequential weak measurements.
Finally, we consider some implications of our results for the problem of distinguishing different causal structures.
\end{abstract}

	
\maketitle


\section{Introduction}

All quantum measurements are subjected to a fundamental trade-off between information gain and disturbance of the measured system. 
In particular, one can perform weak measurements that provide little information but only weakly perturb the system. 
A particularly interesting situation arises when weak measurements are combined with post-selection~\cite{AAV}.
This can be conveniently described within the von Neumann model of quantum measurements, where the quantum system to be measured is coupled via a joint unitary operation to another quantum system, the ``pointer'', which represents the measurement device. 
The measurement is then completed by performing a strong measurement of the pointer.

More formally, consider a system initially prepared (or pre-selected) in a pure state $\ket{\psi}$, and an observable $\hat{A}$ to be weakly measured on it. 
The system-pointer interaction is generated via a Hamiltonian of the form $\hat H = \gamma \hat{A} \otimes \hat{p}$, where $\hat{p}$ denotes the momentum operator acting on the pointer. 
The latter is initially in a state $\ket{\varphi(0)}$, which we shall take here to be a Gaussian wave packet centred at a position $x=0$ with spread (i.e., standard deviation) $\sigma$.
Assuming that we are in the weak measurement regime, with the coupling constant $\gamma$ and interaction time $\Delta t$ such that $g \coloneqq \gamma \Delta t$ is small enough compared to the spread of the pointer, the global state after the coupling is given by
\begin{equation}
e^{-i \hat{H} \Delta t}  \ket{\psi} \ket{\varphi(0)} \approx ( \id -ig\hat{A}\hat{p} )  \ket{\psi} \ket{\varphi(0)} \label{eq:weak_approx1}
\end{equation}
(where tensor products are implicit, and taking $\hbar = 1$). 
For simplicity we will henceforth choose units so that ${g=1}$; 
the strength of the measurement will then be controlled solely by the pointer spread $\sigma$, and the validity of the weak regime will depend only on this being sufficiently large.%
\footnote{\label{fn:g} A characteristic of the weak regime is that an individual measurement of the pointer position following the interaction with the system yields little information, since the pointer spread is much larger than the range of mean pointer positions one obtains (which scales with $g$). As only the ratio between these quantities is important, it is thus sufficient for our analysis to take $g=1$. In the Appendix we present more precise statements of the conditions for the weak regime to be satisfied in the situations we consider throughout this paper.}
Next, the system is post-selected onto the state $\ket{\phi}$ (e.g.\ via a strong projective measurement). 
The final state of the pointer is then (up to normalisation)
\begin{align}
\bra{\phi} (\id -i\hat{A}\hat{p})  \ket{\psi} \ket{\varphi(0)} &= \braket{\phi|\psi} (\id -i  A_\psi^\phi \, \hat{p} ) \ket{\varphi(0)}
\nonumber
\\ & \approx \braket{\phi|\psi} \, e^{-i  A_\psi^\phi \, \hat{p}} \ket{\varphi(0)}, \label{eq:weak_approx2}
\end{align}
where 
\begin{equation}\label{WV}
A_\psi^\phi \coloneqq \frac{\braket{\phi|\hat{A}|\psi} }{\braket{\phi|\psi}}
\end{equation}
is the so-called \emph{weak value} of the observable $\hat{A}$ given the pre-selection in the state $\ket{\psi}$ and post-selection in the state $\ket{\phi}$~\cite{AAV}.
The mean position of the pointer is thus displaced---via the displacement operator $e^{-i A_\psi^\phi \, \hat{p}}$, which generates the (possibly unnormalised) state $\ket{\varphi(A_\psi^\phi)} = e^{-i A_\psi^\phi \, \hat{p}} \ket{\varphi(0)}$; see Appendix---to
\begin{equation}
\braket{\hat{x}} \approx \frac{\bra{\varphi(A_\psi^\phi)} \hat{x} \ket{\varphi(A_\psi^\phi)}}{\braket{\varphi(A_\psi^\phi)|\varphi(A_\psi^\phi)}} = \re(A_\psi^\phi). \label{x_re_WV}
\end{equation}

Notably, the real part of the weak value can become very large when the pre- and post-selected states are almost orthogonal, i.e.\ $|\!\braket{\phi|\psi}\!| \ll 1$. 
In this case, the pointer is, on average, shifted by a large amount. 
Whenever $\re(A_\psi^\phi)$ is not in the interval $[ \lambda_{\text{min}}(\hat{A}), \lambda_{\text{max}}(\hat{A})]$ (where $\lambda_\text{min(max)}(\hat{A})=\min(\max)_k\,\lambda_k(\hat{A})$ and $\lambda_k$ denotes the $k^\text{th}$ eigenvalue of an observable), i.e.\ whenever it is outside of the (convex hull of the) spectrum of $\hat{A}$, the pointer's mean position thus moves beyond where it could have reached under simple weak measurements on an arbitrary pre-selected state without any post-selection.
Indeed, in the absence of post-selection one has (now with exact equalities)
\begin{align}
\braket{\hat{x}} &= \bra{\psi} \bra{\varphi(0)} e^{i \hat{A} \hat{p}} \left( \mathds{1} \otimes \hat{x} \right) e^{-i \hat{A} \hat{p}} \ket{\psi} \ket{\varphi(0)}\notag \\
&= \bra{\psi} \bra{\varphi(0)}  \left( \mathds{1} \otimes \hat{x} + \hat{A} \otimes \mathds{1} \right)  \ket{\psi}
 \ket{\varphi(0)} \notag\\
 &= \bra{\psi} \hat{A} \ket{ \psi} \ \ \in \ [ \lambda_{\text{min}}(\hat{A}), \lambda_{\text{max}}(\hat{A})]. \label{x_no_PS}
\end{align}
Note that $\braket{\hat{x}}$, both with and without post-selection, can be determined experimentally by performing sufficiently many measurements, despite the large variance of the pointer (indeed, to obtain a given accuracy the number of measurements required scales proportionally to $\sigma^2$).

The definition~\eqref{WV} of a weak value can be generalised to post-selections on a given result for any general quantum measurement~\cite{wiseman02,brunner03}. 
In particular, a trivial, deterministic measurement of the identity operator $\id$ amounts to performing no post-selection. 
This allows one to also consider a \emph{weak value with no post-selection}, defined (see Appendix) as%
\footnote{\label{fn:reselection} Note that the weak values $A_{\psi}^{\idmini}$ (in the absence of post-selection) and $A_{\psi}^{\psi}$ (when one post-selects on the initial state, a situation called ``re-selection'' and studied in Ref.~\cite{diosi16}) coincide. However, we emphasise that these correspond to different physical situations; in particular, without post-selection no data is discarded.}
\begin{equation}\label{WV_no_PS}
A_\psi^{\idmini} \coloneqq \braket{\psi|\hat{A}|\psi}.
\end{equation}
With this definition, Eq.~\eqref{x_no_PS} gives $\braket{\hat{x}} = A_\psi^{\idmini} = \re(A_\psi^{\idmini})$: we recover the same relation as in Eq.~\eqref{x_re_WV}, although now $A_\psi^{\idmini}$ is restricted to lie in $[ \lambda_{\text{min}}(\hat{A}), \lambda_{\text{max}}(\hat{A})]$ since here it is simply equal to the expectation value of $\hat{A}$.

The phenomenon of a weak value outside the spectrum of $\hat{A}$ is referred to as an ``anomalous weak value''~\cite{AAV,YV90,weak} since it conflicts with our classical intuition, which would lead us to expect $\braket{\hat{x}}$ to lie within the range of the spectrum of $\hat{A}$. 
This has been observed in many experiments~\cite{ritchie,resch,geoff}, and appears to be directly linked to various (\emph{a priori} unrelated) areas such as tunnelling times~\cite{steinberg} and fast light propagation~\cite{superluminal,solli}. 
In practice, anomalous weak values allow for the detection and precise estimation of very small physical effects~\cite{hosten,dixon,BS,RMP}, via a form of signal amplification,
while they have also helped provide new insights on foundational aspects of quantum mechanics such as contextuality~\cite{pusey14}, counterfactual paradoxes~\cite{aharonov02}, and the nature of the wave function~\cite{lundeen11,kocsis11}.
While astonishing at first sight, anomalous weak values can in fact be intuitively understood in terms of destructive interference of the pointer state, which occurs as a result of post-selection. 
With this in mind and given the rudimentary analysis above, it is rather natural to attribute the origin of anomalous weak values to the presence of post-selection; this opinion indeed seems to be widely shared in the community.  

Here we show, however, that this is not the case in general, and that anomalous weak values can in fact be observed in the absence of post-selection and without discarding any outcomes.
Specifically, we consider a situation in which two successive weak measurements are performed on a quantum system. 
The experiment thus involves two pointers, one associated to each weak measurement, which are measured jointly. 
Considering observables that are simply given by projectors, one expects to find the mean position of each pointer between 0 (the system's state being orthogonal to the projector) and 1 (the system's state being in the range of the projector). 
Yet, we will see that the average of the product of the pointer positions can become negative, something which cannot happen if the measurements are strong or classical. 
This may be understood in terms of the second measurement acting as an effective post-selection of the system, thus creating the desired interference. 
Importantly however, no data is discarded. 

Below, after discussing in detail a simple example of this effect, we turn to the more general scenario of arbitrary sequences of weak measurements without post-selection, deriving bounds on how anomalous a weak value can be obtained in such a scenario and how these relate to the (jointly measured) pointer positions. We finish by discussing the use of anomalous weak values without post-selection to certify particular causal structures between measurements.

\section{Illustrative example}

To start with, let us consider a qubit system initially prepared in the state $\ket{0}$, undergoing a sequence of two weak von Neumann measurements of the projection observables $\ketbra{\psi_j}{\psi_j}$ ($j=1,2$), where the states $\ket{\psi_j}$ and their orthogonal states $\ket{\psi_j^\perp}$ are defined as
\begin{align}
\ket{\psi_j} &= \frac{1}{2} \ket{0} - (-1)^j \frac{\sqrt{3}}{2} \ket{1} , \notag \\
\ket{\psi_j^\perp} &= \frac{\sqrt{3}}{2} \ket{0} + (-1)^j \frac{1}{2} \ket{1}.
\end{align}
To each measurement is associated a pointer in the state $\ket{\varphi_j(x_j)}$, where $x_j$ is the mean position of the pointer wavefunction. 
The two pointers are initially independent, and both centred at $x_j=0$. 
The initial state of the system and pointers is therefore
\begin{align}
\ket{\Psi_0} &= \ket{0}  \ket{\varphi_1(0)}  \ket{\varphi_2(0)}.
\end{align}

Following the von Neumann measurement procedure described earlier with interaction Hamiltonians $\hat H_j = \gamma_j \ketbra{\psi_j}{\psi_j} \hat{p}_j$, the average post-measurement position of the corresponding pointer is (with appropriate units so that $\gamma_j \Delta t_j = 1$ as before) $x_j=1$ if the state of the system is $\ket{\psi_j}$; if the state is $\ket{\psi_j^\perp}$ then the pointer does not move. 
The state of the system and pointers after the interaction with the first pointer is thus
\begin{align}
\ket{\Psi_1} =& \left( \ketbra{\psi_1}{\psi_1}  e^{-i \hat{p}_1} + \ketbra{\psi_1^\perp}{\psi_1^\perp}  \mathds{1}_1 \right) \mathds{1}_2  \,  \ket{\Psi_0} \nonumber \\
=& \frac{1}{2} \ket{\psi_1}  \ket{\varphi_1(1)}  \ket{\varphi_2(0)} \notag\\ 
&+ \frac{\sqrt{3}}{2} \ket{\psi_1^\perp}  \ket{\varphi_1(0)}  \ket{\varphi_2(0)}.\label{stateafterfirstmeasurement}
\end{align}
After interacting with the second pointer, it evolves to
\begin{align}
\ket{\Psi_2} =& \left( \ketbra{\psi_2}{\psi_2}   e^{-i \hat{p}_2} + \ketbra{\psi_2^\perp}{\psi_2^\perp}  \mathds{1}_2 \right) \mathds{1}_1  \, \ket{\Psi_1} \nonumber \\
=& -\frac{1}{4} \ket{\psi_2}  \ket{\varphi_1(1)}  \ket{\varphi_2(1)} \notag\\  
& + \frac{3}{4} \ket{\psi_2}  \ket{\varphi_1(0)}  \ket{\varphi_2(1)} \nonumber \\
&   + \frac{\sqrt{3}}{4} \!\ket{\psi_2^\perp} \! \ket{\varphi_1(1)} \! \ket{\varphi_2(0)} \notag\\
& + \frac{\sqrt{3}}{4} \! \ket{\psi_2^\perp}\! \ket{\varphi_1(0)} \!\ket{\varphi_2(0)}.
\end{align}
Tracing out the system, one finds that the joint pointer state is
\begin{align}\label{jointpointerstate}
\eta_{12} = &\ketbra{\Phi_1^{(1)}}{\Phi_1^{(1)}} \otimes \ketbra{\phi_2(1)}{\phi_2(1)} \nonumber \\
 &+ \ketbra{\Phi_1^{(0)}}{\Phi_1^{(0)}} \otimes \ketbra{\phi_2(0)}{\phi_2(0)},
\end{align}
where
\begin{align}
\ket{\Phi_1^{(1)}} &= \frac14 \big( \ket{\varphi_1(1)} - 3\ket{\varphi_1(0)} \big), \notag \\
\ket{\Phi_1^{(0)}} &= \frac{\sqrt{3}}{4} \big( \ket{\varphi_1(1)} + \ket{\varphi_1(0)} \big) \label{Phi_10}
\end{align}
are (generally unnormalised) states of the first pointer.
The norms of these states, and thus the weight of each state in the mixture $\eta_{12}$, depend on the strength of the first measurement through the overlap $\braket{\varphi_1(1)|\varphi_1(0)}$ of the corresponding pointer states.

Finally the positions of the pointers are measured. 
The quantity of interest is the average of the product of the pointer positions, i.e., the expectation value $\braket{\hat{x}_1 \otimes \hat{x}_2}$. 
From Eq.~\eqref{jointpointerstate}, and using the facts that $\braket{\varphi_2(1)|\hat{x}_2|\varphi_2(1)} = 1$ and $\braket{\varphi_2(0)|\hat{x}_2|\varphi_2(0)} = 0$, we simply find that $\braket{\hat{x}_1 \otimes \hat{x}_2} = \bra{\Phi_1^{(1)}} \hat{x}_1 \ket{\Phi_1^{(1)}}$.

Note that we have not yet specified the strength of either measurement. 
Considering Gaussian pointers with widths $\sigma_j$ for each measurement, we find (see Appendix)
\begin{align}
\braket{\hat{x}_1 \otimes \hat{x}_2} &= \frac{1}{16} \left( 1 - 3e^{-\frac{1}{8 \sigma_1^2}} \right). \label{eq:x1x2_gaussian}
\end{align}
Notice that this quantity depends on $\sigma_1$ but not on $\sigma_2$: the strength of the second measurement has no effect here and can be made as strong as one wishes.%
\footnote{\label{fn:scaling} Note that this means that the variance of $\hat{x}_1\otimes\hat{x}_2$ can be made to be of the order of $\sigma_1^2$ (as opposed to $\sigma_1^2\sigma_2^2$; see Appendix), meaning less statistics are needed to estimate $\braket{\hat{x}_1 \otimes \hat{x}_2}$ to a given precision than would otherwise be the case.}

Since both observables being measured are projectors with spectra $\{ 0,1\}$, one would naturally expect an average value within the range $[0,1]$. 
Regardless of the strength of either measurement, each pointer, taken individually, indeed has an average position in $[0,1]$. Specifically (see Appendix), $\braket{\hat{x}_1} = 1/4$, which is independent of the strength of either measurement, while for the second pointer
$\braket{\hat{x}_2} = \frac{1}{8} \Big(5 - 3e^{-\frac{1}{8 \sigma_1^2}} \Big)$, which ranges from $\braket{\hat{x}_2} \approx 1/4$ when the first measurement is weak ($\sigma_1 \gg 1$), to $\braket{\hat{x}_2} \approx 5/8$ when it is strong ($\sigma_1 \ll 1$).
In the latter regime, Eq.~\eqref{eq:x1x2_gaussian} gives $\braket{\hat{x}_1 \otimes \hat{x}_2} \approx 1/16$, which is consistent with the above argument. However, if the first measurement is sufficiently weak, the average value can become negative; in the limit $\sigma_1 \rightarrow \infty$ we get 
\begin{align}
\braket{\hat{x}_1 \otimes \hat{x}_2} \approx -\frac{1}{8}.
\end{align}

This pointer reading is anomalous in that it gives an average value outside of the natural range of $[0,1]$ that one would expect for the average of a product of two binary 0/1-valued measurements. 
As we will discuss in more detail below, this result can be linked to an anomalous weak value without post-selection, $(\ket{\psi_2}\!\!{\bra{\psi_2}\cdot\ket{\psi_1}}\!\!\bra{\psi_1})_{0}^{\idmini} \coloneqq \braket{0|\psi_2}\!\!\braket{\psi_2|\psi_1}\!\!\braket{\psi_1|0}$ (see Eq.~\eqref{Seq_WV_no_PS} below); specifically, we have here 
\begin{align}\label{eqn:ExampleWV}
\braket{\hat{x}_1 \otimes \hat{x}_2} \approx  \re \big( \braket{0 | \psi_2} \! \braket{\psi_2 | \psi_1} \! \braket{\psi_1 | 0} \big) = -\frac{1}{8} .
\end{align}
We emphasise that this anomalous value is obtained despite the absence of post-selection.
This effect can nevertheless be understood intuitively by considering that the second measurement acts as an effective post-selection on $\ket{\psi_2}$, as the corresponding pointer moves only in this case. 
This becomes apparent upon rewriting the above weak value as 
\begin{align}\label{eqn:missingPSProb}
  \braket{0 | \psi_2} \! \braket{\psi_2 | \psi_1} \! \braket{\psi_1 | 0} =   |\!  \braket{\psi_2 | 0} \! |^2  \frac{\braket{\psi_2 | \psi_1} \! \braket{\psi_1 | 0}}{ \braket{\psi_2 | 0}} ,
\end{align}
which differs from the standard weak value $(\ketbra{\psi_1}{\psi_1})_0^{\psi_2}$ for a post-selection on $\ket{\psi_2}$ only by the factor $|\! \braket{\psi_2 | 0}\!|^2$, which is the probability that the projection of $\ket{0}$ onto $\ket{\psi_2}$ is successful.
As it turns out, this factor ensures in particular that the anomalous weak value without post-selection cannot be arbitrary large, a fact that we prove further below.
For a sequence of two projection observables $\hat{A}$ and $\hat{B}$ (with eigenvalues 0 and 1), the above value of $-1/8$ for the real part is indeed the most anomalous value obtainable (see Appendix).

\section{Analysis for arbitrary observables}

In order to analyse more generally the phenomenon exhibited by the previous example, let us recall some facts about the sequential weak measurement framework within which it can be placed~\cite{Mitchison07}.
In this framework, a sequence of weak measurements is performed, each involving a coupling to a different pointer; these pointers can then be jointly measured following the sequence of interactions and, potentially, a post-selection on the system.
This framework has received much interest recently from both practical~\cite{Piacentini16,Kim18,chen_et_al_18} and foundational~\cite{diosi16,curic18,georgiev18} viewpoints.

For the case of two sequential weak measurements, as in the example of the previous section,
consider thus a system prepared in the pure state $\ket{\psi}$, which is subjected to a sequential weak measurement of the (generally noncommuting) observables $\hat{A}$ then $\hat{B}$, before being post-selected onto the state $\ket{\phi}$.
The system-pointer interaction Hamiltonians are $\hat{H}_1 = \gamma_1 \hat{A} \hat{p}_1$ and $\hat{H}_2=\gamma_2 \hat{B} \hat{p}_2$. 
We will choose again, for simplicity, the coupling constants and interaction times such that $\gamma_j \Delta t_j = 1$, and take Gaussian pointers initially in the states $\ket{\varphi_1(0)}$ and $\ket{\varphi_2(0)}$ with widths $\sigma_1$ and $\sigma_2$, which dictate the measurement strengths.

In analogy to Eq.~\eqref{WV}, the \emph{sequential weak value} $(BA)_\psi^\phi$ is defined,%
\footnote{\label{fn:wv_prod} While this terminology is standard (see, e.g., Refs.~\cite{Resch04,Mitchison07}), note that this should be read as the weak value for measuring $\hat{A}$ then $\hat{B}$ and \emph{not} the weak value of $\hat{B}\hat{A}$, which indeed is not a valid observable in general since it may not be Hermitian. As Eq.~\eqref{eq:WVseq} shows, this quantity nonetheless behaves \emph{as if} it were the weak value of $\hat{B}\hat{A}$.} for a system prepared in $\ket{\psi}$ and post-selected in $\ket{\phi}$, as~\cite{Mitchison07}
\begin{equation}\label{eq:WVseq}
	(BA)_\psi^\phi \coloneqq \frac{\braket{\phi|\hat{B}\hat{A}|\psi}}{\braket{\phi|\psi}}.
\end{equation}
However, while the notion of an anomalous weak value for single (non-sequential) weak measurements is intimately linked to the pointer displacement (and even justified) by the relation $\braket{\hat x}=\re(A_\psi^\phi)$, the relationship between the mean pointer positions and $(BA)_\psi^\phi$ is more subtle for sequential weak measurements. 
In the presence of post-selection, it has instead been shown~\cite{Resch04,Mitchison07} that
\begin{align}\label{eq:SeqWeakMesOutcome}
	\braket{\hat{x}_1\otimes \hat{x}_2}\approx \frac{1}{2}\big(\re[(BA)_\psi^\phi] + \re[ A_\psi^\phi (B_\psi^\phi)^* ]\, \big)
\end{align}
within the weak regime (with large enough widths $\sigma_1$ and $\sigma_2$). 
This cautions that some care must be taken when linking (possibly anomalous) pointer positions to weak values.

The sequential weak value of Eq.~\eqref{WV_no_PS} can be generalised to the case without post-selection, by defining, in a similar way to before, the \emph{sequential weak value with no post-selection} as
\begin{equation}\label{Seq_WV_no_PS}
(BA)_\psi^{\idmini} \coloneqq \braket{\psi|\hat{B}\hat{A}|\psi}.
\end{equation}
Indeed, this quantity has previously been considered in the study of time asymmetry in sequential weak measurements~\cite{bednorz13,curic18} and their quasiprobabilistic interpretation~\cite{bednorz10}.
Connecting this to the pointer positions, we prove in the Appendix that, contrary to Eq.~\eqref{eq:SeqWeakMesOutcome} (which was obtained with post-selection), we recover here the direct relation%
\footnote{As in the single measurement case, the weak values $(BA)_{\psi}^{\idmini}$ (in the absence of post-selection) and $(BA)_{\psi}^{\psi}$ (in the case of ``re-selection''~\cite{diosi16}) coincide; cf.\ Footnote~\ref{fn:reselection}.
Recall, however, that these correspond to different physical situations; crucially here, the mean pointer positions in Eqs.~\eqref{eq:SeqWeakMesOutcomeNoPS} and~\eqref{eq:SeqWeakMesOutcome} differ in general between these scenarios. See Appendix for further discussion.\label{fn:reselection2}}
\begin{equation}\label{eq:SeqWeakMesOutcomeNoPS}
	\braket{\hat{x}_1\otimes \hat{x}_2}\approx \re[(BA)_\psi^{\idmini}],
\end{equation}
as anticipated already in Eq.~\eqref{eqn:ExampleWV}, which holds as long as the \emph{first} measurement is sufficiently weak~\cite{bednorz10}. 

This justifies that our earlier illustrative example could indeed be interpreted as yielding an anomalous weak value without post-selection.
Crucially, although for a single measurement without post-selection $A_\psi^{\idmini}$ is simply the expectation value of $\hat{A}$, no such interpretation can be given to $(BA)_\psi^{\idmini}$:
it is the weak value of a sequence of measurements and simply behaves \emph{as if} it were the weak value of the operator $\hat{B}\hat{A}$ which, as already mentioned, is only Hermitian -- and thus defines an observable -- if $\hat{A}$ and $\hat{B}$ commute (see also footnote~\ref{fn:wv_prod}).
In particular, this implies that unless $\hat{A}$ and $\hat{B}$ commute, $(BA)_\psi^{\idmini}$ need not be contained within the interval $[\Lambda_\text{min}(\hat{A},\hat{B}),\Lambda_\text{max}(\hat{A},\hat{B})]$, where $\Lambda_\text{min(max)}(\hat{A},\hat{B})=\min(\max)_{k,\ell}\,\lambda_k(\hat{A})\lambda_\ell(\hat{B})$, as one one would naturally expect for the product of outcomes for a measurement of $\hat{A}$ then $\hat{B}$~\cite{Lundeen12,Thekkadath16}.

Nevertheless, as we noted after Eq.~\eqref{eqn:missingPSProb}, the value of $(BA)_\psi^{\idmini}$ cannot be amplified arbitrarily. It is possible to place a more quantitive bound on the values that it can in fact take.
Using the Cauchy-Schwartz inequality, we indeed have
\begin{align}\label{eqn:ampliBound}
	|(BA)_\psi^{\idmini}| & = |\braket{\psi|\hat{B}\hat{A}|\psi}| \notag \\
	& \le \sqrt{\braket{\psi|\hat{A}^2|\psi}\braket{\psi|\hat{B}^2|\psi}} \le \lVert\hat A\rVert \, \lVert\hat B\rVert,
\end{align}
(where $\lVert\cdot\rVert$ is the spectral norm).
Thus, although the mean pointer position can show anomalous weak values without post-selection, the \emph{magnitude} of the mean pointer position cannot be pushed outside what one can obtain using strong measurements.

The bound above implies in particular that for observables with symmetric spectra (with respect to 0), the real part of the weak value -- and therefore the mean product of pointer positions, see Eq.~\eqref{eq:SeqWeakMesOutcomeNoPS} -- cannot be anomalous; anomalous pointer positions are only obtained for observables with asymmetric spectra, such as projection observables.
Nevertheless, one can also obtain complex weak values for observables with symmetric spectra.
Take, for example, a system initially prepared in the $(+1)$-eigenstate $\ket{0}$ of the Pauli matrix $\hat\sigma_\textsc{z}$, on which a sequential weak measurement of the Pauli observables $\hat\sigma_\textsc{y}$ and $\hat\sigma_\textsc{x}$ is performed.
One thus obtains $(\sigma_\textsc{x}\sigma_\textsc{y})_{0}^{\idmini}=i$.
The imaginary part of the weak value here can be detected by measuring the pointer momenta~\cite{Resch04,RMP} (see Appendix).
Such complex anomalous weak values cannot be obtained without post-selection with only a single weak measurement or a sequence of weak measurements of commuting observables, and can thus themselves be considered anomalous in this sense.

\section{More measurements}

Eq.~\eqref{eqn:ampliBound} might bound how anomalous a weak value can be without post-selection, but it is not generally tight.
For two projection observables $\hat A$ and $\hat B$ (with eigenvalues $\pm 1$), for example, it only implies a bound $\re[(BA)_\psi^{\idmini}] \ge -1$; nevertheless, as we prove in the Appendix, the value of $-1/8$ obtained earlier for the real part of the weak value is the most negative value that one can obtain.
Can one do better by considering longer sequences of successive weak measurements?
Here we will see that this question has a subtle answer: the weak value itself can approach $-1$, but this will not mean the average product of the pointer positions does so as well.

For a sequence of $n$ observables $\hat{A}_1,\dots,\hat{A}_n$ to be measured weakly on the state $\ket{\psi}$ before a post-selection on $\ket{\phi}$, the sequential weak value is defined (following, e.g., Ref.~\cite{Mitchison07}) as 
\begin{equation}\label{eq:weakvaluen}
	(A_n\cdots A_1)_\psi^\phi \coloneqq \frac{\braket{\phi|\hat{A}_n \cdots \hat{A}_1|\psi}}{\braket{\phi|\psi}}.
\end{equation}
When no post-selection is performed, this can be generalised to
\begin{equation}\label{eq:weakvaluen_no_PS}
	(A_n\cdots A_1)_\psi^{\idmini} \coloneqq \braket{\psi|\hat{A}_n \cdots \hat{A}_1|\psi},
\end{equation}
in analogy to the cases discussed earlier.
As we show in the Appendix, a similar bound to Eq.~\eqref{eqn:ampliBound} can be derived, namely
\begin{align}\label{eqn:ampliBoundNobs}
	|(A_n\cdots A_1)_\psi^{\idmini}| \ \le \ \prod_{j=1}^n \lVert\hat{A}_j\rVert \, .
\end{align}

For $n$ projection observables, this implies the bound $\re[(A_n\cdots A_1)_\psi^{\idmini}] \ge -1$. 
As it turns out, it is possible to obtain an anomalous sequential weak value without post-selection approaching $-1$ and thus saturating this bound in the limit $n\to\infty$.
To see this, take the initial state of the system to be $\ket{\psi}=\ket{0}$ and consider the sequence of $n$ qubit projectors $\hat{A}_j=\ketbra{a_j}{a_j}$ with $\ket{a_j}=\cos(\frac{j\pi}{n+1})\ket{0}+\sin(\frac{j\pi}{n+1})\ket{1}$ for $j=1,\dots,n$.
This sequence of weak measurements gives 
\begin{equation}
	(A_n\cdots A_1)_\psi^{\idmini} = -\left(\cos\tfrac{\pi}{n+1}\right)^{n+1} \xrightarrow{\quad n\to \infty \quad} -1.
\end{equation}
Note that for $n=2$ this coincides precisely with the explicit two-measurement example we began with.

As discussed above, for two sequential weak measurements in the absence of post-selection, the mean product of the pointer positions gives precisely the real part of the sequential weak value; see Eq.~\eqref{eq:SeqWeakMesOutcomeNoPS}.
However, for $n>2$ measurements this direct relationship is broken and the mean product of the pointer positions corresponds instead to a mixture of sequential weak values for $2^{n-2}$ different permutations of the observables (see the Appendix for an explicit expression).
For example, for $n=3$, in the weak regime, we have~\cite{diosi16,curic18}
\begin{align}\label{eq:Pointer3measurements}
	\braket{\hat{x}_1\otimes \hat{x}_2 \otimes \hat{x}_3} \approx \frac{1}{2}\big( & \re[(A_3A_2A_1)_\psi^{\idmini}] \notag\\
	& + \re[(A_2A_3A_1)_\psi^{\idmini}]\big).
\end{align}
The real part of $(A_3A_2A_1)_\psi^{\idmini}$ is thus not directly observed.
However, as we show in the Appendix, its value (as well as the imaginary part) can nonetheless be deduced experimentally by measuring several different expectation values of the products of pointer positions and momenta~\cite{Mitchison07}. 

Interestingly, by numerically minimising the mean product of the pointer positions for sequences of up to 5 projection observables, we were unable to obtain a value smaller than $-1/8$, and we conjecture that this is in fact the case for all $n$.
Thus, although the weak value itself can be brought arbitrarily close to $-1$, it seems that additional sequential weak measurements may not lead to ``more anomalous'' pointer positions.
This behaviour highlights oft-overlooked subtleties in the connection between anomalous weak values and pointer positions: for individual weak measurements, there is a direct correspondence between the pointer position and (the real part of) the weak value, and an anomalous weak value has an immediate physical relevance.
For sequential weak measurements, a distinction must be made between anomalous weak values and anomalous pointer positions (with post-selection, this is already the case for two measurements; see Eq.~\eqref{eq:SeqWeakMesOutcome} or Ref.~\cite{Mitchison07}).
It is the latter of these phenomena that arguably provides the more important anomaly, leading to average measurement results that lie outside the range of values one would expect to be attainable.

In contrast, this divergence between weak values and pointer positions for sequential weak measurements means that, in general, it is more difficult to give a clear physical interpretation to sequential weak values, anomalous or not.
Indeed, while some authors have argued that weak values for single weak measurements should be considered real properties of quantum states with direct physical meaning~\cite{Vaidman96,Vaidman17}, it is unclear whether such arguments are justified for sequential weak values~\cite{Mitchison07} given the lack of examples of physical scenarios where they play a direct, crucial role.

\section{Further Discussion}

Compare the situation of a sequential weak measurement of two observables $\hat{A}$ and $\hat{B}$ with the alternative in which a bipartite system $\ket{\psi_{ab}}\in \mathcal{H}_a\otimes\mathcal{H}_b$ is prepared and $\hat{A}$ and $\hat{B}$ are weakly measured on the two different substituent systems.
One can view this either as a measurement of the joint observable $\hat{A}\otimes \hat{B}$ (with two different pointers, one coupled to each observable) or a sequential measurement of the commuting observables $\hat{A}\otimes \id$ and $\id \otimes \hat{B}$.
In the absence of any post-selection, one has $(A\otimes B)_{\psi_{ab}}^{\idmini} = \braket{\psi_{ab}|\hat{A}\otimes \hat{B}|\psi_{ab}}$ which, being just an expectation value, cannot lie outside the spectrum of the product observable $\hat{A} \otimes \hat{B}$.
For tensor product measurements, an anomalous weak value is thus unobtainable without post-selection.

This observation raises some interesting implications.
Consider for example a scenario in which two parties, Alice and Bob, each operate in a closed laboratory.
Each receives a system, performs a weak measurement, and sends the resulting system out;
they then come together to jointly measure their pointers.
By repeating this many times (or on a large number of systems), they thus determine $\braket{\hat{x}_1\otimes \hat{x}_2}$.
If Alice and Bob have no knowledge of their causal relationship, they could unknowingly be weakly measuring the same system at different times (either Alice then Bob, or vice versa), or measuring different parts of a (potentially entangled) bipartite system.
By observing an anomalous weak value they can differentiate between these two scenarios.

The problem of distinguishing these two causal structures for quantum systems -- the former is known as a \emph{direct cause} relationship, while the later a \emph{common cause} relationship, since any correlations must be due to a (possibly quantum) common cause -- has been the subject of recent interest; see, e.g., Refs.~\cite{fitzsimons15,ried15,costa16,allen17,giarmatzi18}. 
An anomalous weak value thus provides a novel way to witness a direct causal relationship and distinguish between these cases.
We leave it as an open question whether such a witness can be found whenever Alice and Bob are capable of signalling to each other; i.e., if whenever they are connected by a quantum channel of nonzero capacity they can always find a pair of observables to measure that would generate an anomalous weak value without post-selection.

Recently, there has also been substantial interest in quantum processes that are not consistent with any definite (possibly stochastic) causal ordering~\cite{oreshkov12}, and practical approaches to witness such ``indefinite'' causal orders have been developed~\cite{araujo15,branciard16} and experimentally tested~\cite{rubino17,goswami18}.
It would be interesting to see whether indefinite causal orders may also be witnessed by, for example, producing anomalous weak values or pointer positions lying further outside the expected range than possible in a well-defined causal structure.

The ability for sequential weak measurements to lead to anomalous measurement results even in the absence of post-selection is thus interesting not only in clarifying the relation between post-selection and anomalous weak values, but also in that it raises interesting foundational questions. This sequential weak measurement scheme has, for example, already been used to study time asymmetry in quantum measurements~\cite{bednorz13,curic18,franke12} and the role of anomalous weak values in this context would be interesting to study further. One may expect anomalous weak values without post-selection -- just as for those with post-selection -- to shed light on other foundational results; for example, it would be interesting to see whether they can lead to proofs of contextuality as they do in the standard single weak measurement scenario~\cite{pusey14}.

\paragraph{Acknowledgements}

We thank Yelena Guryanova for fruitful discussions. We acknowledge financial support from the `Retour Post-Doctorants' (ANR-13-PDOC-0026) and the `Investissements d'avenir' (ANR-15-IDEX-02) programs of the French National Research Agency, and from the Swiss national science foundation (Starting grant DIAQ [grant No.\ 200021\_169002], grant No.\ 200020\_165843, and NCCR-QSIT).

\nocite{apsrev41Control} 
\bibliographystyle{apsrev4-1_modified}
\bibliography{weakvalues}

\vspace{1cm}

\onecolumngrid

\appendix

\renewcommand{\theequation}{A\arabic{equation}}
\setcounter{equation}{0}

\renewcommand{\thetable}{A\arabic{table}}
\setcounter{table}{0}

\renewcommand{\thefigure}{A\arabic{figure}}
\setcounter{figure}{0}

\section{Weak measurements with Gaussian pointers}

The weak measurement regime can in general be defined for any type of von Neumann measurement interaction scheme, as introduced in the main text, by comparing the various parameters that describe it: the strength of the measurement interaction ($\gamma$), the time of the interaction ($\Delta t$), the width of the measurement pointer ($\sigma$), the eigenspectrum of the observable $\hat{A}$ being measured as well as the weak values to be considered. The relation between weak values and pointer positions can then be obtained by taking the appropriate limits.

As the point of our paper is to analyse specific cases of anomalous weak values and anomalous pointer positions, for simplicity we choose a specific form for the pointer states, namely, Gaussian states.
By a Gaussian pointer, we mean a pointer whose state $\ket{\varphi(a)}$ is described by a Gaussian wavefunction as follows:
\begin{align}\label{eq:gaussianpointer}
	\ket{\varphi(a)} & \coloneqq \int_{-\infty}^{\infty} \mathrm{d} x \ \left( \frac{1}{2\pi\sigma^2} \right)^{1/4} \exp\left[ -\frac{(x-a)^2}{4\sigma^2} \right] \ket{x},
\end{align}
where $\{\ket{x}\}_x$ is a continuous eigenbasis of the pointer position $\hat{x}$.
For $a \in \mathbb{R}$, $\ket{\varphi(a)}$ is properly normalised; for a complex value of $a$, its norm is $e^{\frac{\im(a)^2}{4\sigma^2}}$.
The mean position of the pointer in the state $\ket{\varphi(a)}$ (possibly after renormalisation) is $\braket{\hat x} = \re(a)$ and its variance is $\braket{\hat x^2} - \braket{\hat x}^2 = \sigma^2$, while the mean value of the momentum operator $\hat p = -i \frac{\partial}{\partial x}$ (taking $\hbar = 1$) is $\braket{\hat p} = \frac{\im(a)}{2\sigma^2}$ ($=0$ for $a \in \mathbb{R}$).
Prior to the measurement, we always start with $a=0$. Note that an operator of the form $e^{-i \alpha \hat p}$ (for any $\alpha \in \mathbb{C}$) acts as a displacement operator, such that $e^{-i \alpha \hat p}\ket{\varphi(a)} = \ket{\varphi(a+\alpha)}$.

\bigskip

Let us clarify here the conditions that define the standard weak measurement regime, under which the approximations of Eqs.~\eqref{eq:weak_approx1}--\eqref{eq:weak_approx2} of the main text are valid.
As everywhere in the paper, we choose units such that $g = \gamma \Delta t = 1$; as we noted in Footnote~\ref{fn:g}, this effectively fixes the scale for the pointer movement, while the weak regime in general depends on the spread of the pointer with respect to this scale, i.e., on $\sigma/g$.
By considering the spectral decomposition $\hat A = \sum_k a_k \ketbra{a_k}{a_k}$ of the observable under consideration and the completeness relation $\id = \sum_k \ketbra{a_k}{a_k}$ (with $\{\ket{a_k}\}_k$ an orthonormal basis of the system Hilbert space), one can write $e^{-i \hat A \otimes \hat p} = \sum_k \ketbra{a_k}{a_k} \otimes e^{-i a_k \hat p}$ and $\id - i \hat{A} \otimes \hat{p} = \sum_k \ketbra{a_k}{a_k} \otimes ( \id - i a_k \hat p)$, so that the difference between the left and right hand sides of Eq.~\eqref{eq:weak_approx1} is
\begin{align}
\sum_k \braket{a_k|\psi} \ket{a_k} \otimes \ket{\delta_k} \quad \text{with} \quad \ket{\delta_k} \coloneqq e^{-i a_k \hat p} \ket{\varphi(0)} & - ( \id - i a_k \hat p) \ket{\varphi(0)} = \ket{\varphi(a_k)} - ( \id - i a_k \hat p) \ket{\varphi(0)}.
\end{align}
The approximation of Eq.~\eqref{eq:weak_approx1} is valid if for each $k$ (for which $|\!\braket{a_k|\psi}\!|$ is non-negligible), the norm of $\ket{\delta_k}$ is small enough (compared e.g. to that of the lhs of Eq.~\eqref{eq:weak_approx1}, which is 1).
Using Eq.~\eqref{eq:gaussianpointer}, one finds
\begin{align}
\braket{\delta_k|\delta_k} = 2 \Big( 1 - e^{- \frac{a_k^2}{8\sigma^2}} \Big) + \frac14\frac{a_k^2}{\sigma^2} \Big( 1 - 2 e^{- \frac{a_k^2}{8\sigma^2}} \Big) = \frac{3}{64} \Big( \frac{a_k}{\sigma} \Big)^4 + O\Big[ \Big( \frac{a_k}{\sigma} \Big)^6 \Big],
\end{align}
which is indeed small if
\begin{align}
\sigma \gg |a_k| \quad \forall \, k. \label{eq:weak_regime_cond1}
\end{align}
Similarly, the difference between the two lines of Eq.~\eqref{eq:weak_approx2} is (ignoring the common prefactor $\braket{\phi|\psi}$)
\begin{align}
e^{-i \, A_\psi^\phi \, \hat{p}} \ket{\varphi(0)} - (\id -i \, A_\psi^\phi \, \hat{p} ) \ket{\varphi(0)} = \ket{\varphi(A_\psi^\phi)} - (\id -i \, A_\psi^\phi \, \hat{p} ) \ket{\varphi(0)},
\end{align}
which is also small if
\begin{align}
\sigma \gg |A_\psi^\phi|. \label{eq:weak_regime_cond2}
\end{align}

Thus the weak regime is valid whenever the two conditions~\eqref{eq:weak_regime_cond1} and~\eqref{eq:weak_regime_cond2} are fulfilled.

\section{Relating weak values to pointer positions and momenta}

\subsection{The case of a single weak measurement}

It will be useful to relate here the formula for a weak value to the pointer position and momentum in a more general setting than that considered in the main text, where the initial state is not considered \emph{a priori} to be pure, and where post-selection is conditioned on a given result of an arbitrary Positive-Operator Valued Measure (POVM) measurement on the system, rather than a projective measurement.
In such a setting, Eq.~\eqref{WV} can be generalised to
\begin{equation}\label{WV_general}
A_{\rho}^{E} \coloneqq \frac{\Tr (E\hat{A}\rho)}{\Tr (E\rho)},
\end{equation}
which now defines the weak value of the observable $\hat{A}$, given the pre-selection in the state $\rho$ and post-selection by the POVM element $E$. 
This definition was first proposed explicitly in Ref.~\cite{wiseman02} (although earlier alluded to in Ref.~\cite{AV02}), and shown to be indeed the natural generalisation of the standard definition~\eqref{WV} of a weak value. 
Note already that this definition indeed reduces to Eq.~\eqref{WV} if the preparation is a pure state $\ket{\psi}$ (i.e.\ for $\rho = \ketbra{\psi}{\psi}$) and the post-selection is a projection onto another pure state $\ket{\phi}$ (for $E = \ketbra{\phi}{\phi}$). It also allows one to define a weak value with no post-selection by taking a trivial POVM element $E = \id$, which indeed reduces to the definition of Eq.~\eqref{WV_no_PS} in the case of a pure state $\rho = \ketbra{\psi}{\psi}$ -- and which simply coincides here, in the case of a single observable, with the expectation value of $\hat A$; note in particular that, contrary to a general weak value, the weak value with no post-selection is linear in the pre-selected state.

\bigskip

In the von Neumann measurement scenario that we consider here, one thus prepares the density matrix $\rho$, weakly measures $\hat A \, (= \sum_k a_k \ketbra{a_k}{a_k})$ (with a pointer in a Gaussian state as described above), and finally post-selects an outcome corresponding to the POVM element $E$. The initial density matrix of the system and pointer is given by
\begin{align}
	\varrho  &= \rho \otimes \ketbra{\varphi(0)}{\varphi(0)} = \sum_{k\ell} \ketbra{a_k}{a_k} \rho \ketbra{a_\ell}{a_\ell} \otimes \ketbra{\varphi(0)}{\varphi(0)}.
\end{align}
Under the interaction between the system and the pointer, the joint state evolves to
\begin{align}
e^{-i\hat{A}\hat{p}} \varrho \, e^{+i\hat{A}\hat{p}} &= \sum_{k\ell} \ketbra{a_k}{a_k} \rho \ketbra{a_\ell}{a_\ell} \otimes e^{-i a_k \hat{p}} \ketbra{\varphi(0)}{\varphi(0)} e^{+i a_\ell \hat{p}} \\
&= \sum_{k\ell} \ketbra{a_k}{a_k} \rho \ketbra{a_\ell}{a_\ell} \otimes \ketbra{\varphi(a_k)}{\varphi(a_\ell)}.
\end{align}
Due to the post-selection upon $E$, the state of the pointer is then projected onto the (unnormalised) state
\begin{align}
\eta = \Tr_S \left( (E \otimes \id) \, \big( e^{-i\hat{A}\hat{p}} \varrho \, e^{+i\hat{A}\hat{p}} \big) \right) &= \sum_{k\ell} \Tr \Big( E \ketbra{a_k}{a_k} \rho \ketbra{a_\ell}{a_\ell} \Big) \ketbra{\varphi(a_k)}{\varphi(a_\ell)}
\end{align}
(where $\Tr_S$ is the partial trace over the state of the system).

The expectation value of the position of the pointer, given that the post-selection was successful, is
\begin{align}
	\braket{\hat{x}} = \frac{\Tr \left( \hat{x}\, \eta \right) }{\Tr \left( \eta \right) } &= \frac{ \sum_{k\ell} \Tr \Big( E \ketbra{a_k}{a_k} \rho \ketbra{a_\ell}{a_\ell} \Big) \braket{\varphi(a_\ell)| \hat{x} | \varphi(a_k) } }{ \sum_{mn} \Tr \Big( E \ketbra{a_m}{a_m} \rho \ketbra{a_n}{a_n} \Big) \braket{\varphi(a_n)| \varphi(a_m) } }.
\end{align}
Evaluating the expressions in the fraction above for the Gaussian pointer of Eq.~\eqref{eq:gaussianpointer} (with $a_k, a_\ell, a_m, a_n \in \mathbb{R}$), and taking the weak regime approximation in which $\sigma \gg |a_k-a_\ell|$, $\sigma \gg |a_m-a_n|$, one finds, to the lowest order,
\begin{align}
	\braket{\varphi(a_\ell)| \hat{x} | \varphi(a_k) } &= \frac{a_k+a_\ell}{2} e^{-\frac{(a_k-a_\ell)^2}{8 \sigma^2}} \approx \frac{a_k+a_\ell}{2} , \qquad \braket{\varphi(a_n)| \varphi(a_m) } = e^{-\frac{(a_m-a_n)^2}{8 \sigma^2}} \approx 1 , \label{eq:gaussian_approx}
\end{align}
so that we are left with
\begin{align}
	\braket{\hat{x}} &\approx \frac{ \sum_{k\ell} \Tr \Big( E \ketbra{a_k}{a_k} \rho \ketbra{a_\ell}{a_\ell} \Big) (a_k+a_\ell)/2 }{ \sum_{mn} \Tr \Big( E \ketbra{a_m}{a_m} \rho \ketbra{a_n}{a_n} \Big) } = \frac{ \big( \Tr (E\hat{A}\rho) + \Tr(E\rho \hat{A}) \big)/2 }{ \Tr (E\rho) } = \re \left( \frac{\Tr (E\hat{A}\rho)}{\Tr (E\rho)} \right),
\end{align}
where we used the spectral decomposition of $\hat{A}$ and the cyclic property of the trace, together with the fact that $E$, $\hat{A}$ and $\rho$ are Hermitian and thus $\Tr(EP\hat{A})=\Tr((EP\hat{A})^\dagger)^*$.
Recalling the generalised definition~\eqref{WV_general} of a weak value, we thus find that the mean position of the pointer (when the post-selection is successful) is, as in Eq.~\eqref{x_re_WV},
\begin{equation}
\braket{\hat x} \approx \re (A_{\rho}^{E}). \label{x_re_WV_gen}
\end{equation}

Similarly, by using the fact that
\begin{equation}\label{eq:gaussianx2}
	\braket{\varphi(a_\ell)| \hat{x}^2 | \varphi(a_k) } = \left( \sigma^2 + \Big(\frac{a_k+a_\ell}{2}\Big)^2 \right) e^{-\frac{(a_k-a_\ell)^2}{8 \sigma^2}} \approx \sigma^2 + \Big(\frac{a_k+a_\ell}{2}\Big)^2
\end{equation}
one may verify that the variance of the pointer position is
\begin{equation}
	(\Delta \hat{x})^2=\braket{\hat{x}^2}-\braket{\hat{x}}^2 \approx \sigma^2 + \frac{1}{2}\left[ \re\!\big((A^2)_{\rho}^{E}\big) + \frac{\Tr(E\hat{A}\rho \hat{A})}{\Tr(E\rho)} \right] - \big[\re(A^E_{\rho})\big]^2,
\end{equation}
which, for a weak enough measurement is dominated by $\sigma^2$.
Note that for pure pre- and post-selected states (i.e.\ for $\rho = \ketbra{\psi}{\psi}$ and $E = \ketbra{\phi}{\phi}$) one has $\frac{\Tr(E\hat{A}\rho \hat{A})}{\Tr(E\rho)} = \frac{\braket{\phi|\hat{A}|\psi}\!\braket{\psi|\hat{A}|\phi}}{\braket{\phi|\psi}\!\braket{\psi|\phi}}=|A_{\psi}^{\phi}|^2$.
In the absence of post-selection (i.e., $E=\id$) on the other hand, when the weak value is equal to the expectation value, one has simply $(\Delta \hat{x})^2 = \sigma^2 + (\Delta \hat{A})^2$ (where $(\Delta \hat{A})^2$ is the variance of the observable $\hat{A}$ for the state $\rho$).

\bigskip

One may also consider measuring the expectation value of the \emph{momentum} of the pointer instead, conditioned again on a successful post-selection:
\begin{align}
	\braket{\hat{p}} = \frac{\Tr \left( \hat{p}\, \eta \right) }{\Tr \left( \eta \right) } &= \frac{ \sum_{k\ell} \Tr \Big( E \ketbra{a_k}{a_k} \rho \ketbra{a_\ell}{a_\ell} \Big) \braket{\varphi(a_\ell)| \hat{p} | \varphi(a_k) } }{ \sum_{mn} \Tr \Big( E \ketbra{a_m}{a_m} \rho \ketbra{a_n}{a_n} \Big) \braket{\varphi(a_n)| \varphi(a_m) } }.
\end{align}
The relevant quantity for the pointer states is, in the weak regime approximation,
\begin{align}\label{eq:gaussian_approxP}
	\braket{\varphi(a_\ell)| \hat{p} | \varphi(a_k) } &= \frac{1}{2\sigma^2} \, \frac{a_k - a_\ell}{2i} e^{-\frac{(a_k-a_\ell)^2}{8 \sigma^2}} \approx \frac{1}{2\sigma^2} \, \frac{a_k - a_\ell}{2i},
\end{align}
from which (together with Eq.~\eqref{eq:gaussian_approx}) we find that the expectation of the pointer's momentum is
\begin{align}
	\braket{\hat{p}} &\approx \frac{1}{2\sigma^2} \frac{ \sum_{k\ell} \Tr \Big( E \ketbra{a_k}{a_k} \rho \ketbra{a_\ell}{a_\ell} \Big) (a_k-a_\ell)/2i }{ \sum_{mn} \Tr \Big( E \ketbra{a_m}{a_m} \rho \ketbra{a_n}{a_n} \Big) } = \frac{1}{2\sigma^2} \frac{ \big( \Tr (E\hat{A}\rho) - \Tr(E\rho \hat{A}) \big)/2i }{ \Tr (E\rho) } = \frac{1}{2\sigma^2} \im \left( \frac{\Tr (E\hat{A}\rho)}{\Tr (E\rho)} \right),
\end{align}
that is,
\begin{equation}
\braket{\hat p} \approx \frac{1}{2\sigma^2} \im (A_{\rho}^{E}). \label{p_im_WV_gen}
\end{equation}

The expectation value of the momentum is thus directly linked here to the imaginary part of the weak value. Using Eqs.~\eqref{x_re_WV_gen} and~\eqref{p_im_WV_gen}, one may therefore recover both the real and imaginary parts of $A_{\rho}^{E}$ from the expectation values of the pointer's position and momentum (in a regime where $\sigma$ is large enough to ignore higher order terms, but not so small as to render the term that remains above, with the pre-factor $\frac{1}{2\sigma^2}$, unmeasurable).

Note that, unlike the expression for $\braket{\hat{x}}$, the above expression for $\braket{\hat{p}}$ depends explicitly on the width of the pointer.
However, by using the fact that
\begin{equation}\label{eq:gaussianp2}
	\braket{\varphi(a_\ell)| \hat{p}^2 | \varphi(a_k) } = \frac{1}{4\sigma^4} \left( \sigma^2 - \Big(\frac{a_k-a_\ell}{2}\Big)^2 \right) e^{-\frac{(a_k-a_\ell)^2}{8 \sigma^2}} \approx \frac{1}{4\sigma^4} \left( \sigma^2 - \Big(\frac{a_k-a_\ell}{2}\Big)^2 \right),
\end{equation}
the error in the pointer momentum is seen to scale similarly, and so the relative error scales in the same way as that of the pointer position. 
Indeed, one finds
\begin{equation}
	(\Delta\hat{p})^2=\braket{\hat{p}^2}-\braket{\hat{p}}^2 \approx\frac{1}{4\sigma^4}\left(\sigma^2 - \frac{1}{2}\left[ \re\!\big((A^2)_{\rho}^{E}\big) - \frac{\Tr(E\hat{A}\rho \hat{A})}{\Tr(E\rho)}\right] - \big[\im(A^E_{\rho})\big]^2 \right).
\end{equation}

\subsection{Two sequential weak measurements}

Let us now turn to the sequential measurement of two observables $\hat A$ and $\hat B$. The sequential weak value $(BA)_\psi^\phi$ for a system prepared in the pure state $\ket{\psi}$ and post-selected in $\ket{\phi}$, defined in Eq.~\eqref{eq:WVseq}, was introduced in Ref.~\cite{Mitchison07}.
As in the previous section, one may consider a more general setting where the initial state $\rho$ is not necessarily assumed to be pure, and the post-selection is conditioned on a general POVM element $E$. In such a case, the definition of Eq.~\eqref{eq:WVseq} can, similarly to Eq.~\eqref{WV_general}, naturally be generalised to 
\begin{equation}\label{eq:WVseq_general}
	(BA)_\rho^E \coloneqq \frac{\Tr (E\hat{B}\hat{A}\rho)}{\Tr (E\rho)}.
\end{equation}
One can indeed verify that one recovers Eq.~\eqref{eq:WVseq} for $\rho = \ketbra{\psi}{\psi}$ and $E = \ketbra{\phi}{\phi}$. As before, this definition also allows one to define a sequential weak value with no post-selection by taking the trivial POVM element $E = \id$, as in Eq.~\eqref{Seq_WV_no_PS} for the case of a pure state $\rho = \ketbra{\psi}{\psi}$. 
Again, and contrary to a general sequential weak value, the sequential weak value with no post-selection is linear in the pre-selected state; note, however, that it no longer coincides with an expectation value, as in general the product $\hat{B}\hat{A}$ is not Hermitian, and thus does not define a valid observable.

\bigskip

We consider here a sequential von Neumann measurement scenario where two separate Gaussian pointers (labelled by the subscripts $j=1,2$) are used to measure $\hat A \, (= \sum_k a_k \ketbra{a_k}{a_k})$ and $\hat B \, (= \sum_m b_m \ketbra{b_m}{b_m})$ on a system prepared in the state $\rho$ and post-selected on a POVM element $E$.
Similarly to the analysis in the previous section, the final (unnormalised) state of the two pointers after the post-selection is given by (with implicit identity operators)
\begin{align}\label{eq:eta}
\eta &= \Tr_S \left( E \, e^{-i\hat{B}\hat{p}_2} e^{-i\hat{A}\hat{p}_1} \big( \rho \otimes \ketbra{\varphi_1(0)}{\varphi_1(0)} \otimes \ketbra{\varphi_2(0)}{\varphi_2(0)} \big)  e^{+i\hat{A}\hat{p}_1} e^{+i\hat{A}\hat{p}_2} \right) \notag \\
&= \sum_{k\ell mn} \Tr \big( E \, \ket{b_m}\!\!\braket{b_m|a_k}\!\!\bra{a_k} \rho \ket{a_\ell }\!\!\braket{a_\ell |b_n} \!\!\bra{b_n} \big) \ketbra{\varphi_1(a_k)}{\varphi_1(a_\ell )} \otimes \ketbra{\varphi_2(b_m)}{\varphi_2(b_n)}.
\end{align}

Using the weak regime approximations of Eq.~\eqref{eq:gaussian_approx}, for both the weak measurement of $\hat{A}$ and of $\hat{B}$, we find that the expectation value of the product of the pointer positions, given that the post-selection was successful, is
\begin{align}
	\braket{\hat{x}_1 \otimes \hat{x}_2} = \frac{\Tr \left( \hat{x}_1 \otimes \hat{x}_2\, \eta \right) }{\Tr \left( \eta \right) } &\approx \frac{ \sum_{k\ell mn} \Tr \big( E \, \ket{b_m}\!\!\braket{b_m|a_k}\!\!\bra{a_k} \rho \ket{a_\ell}\!\!\braket{a_\ell|b_n} \!\!\bra{b_n} \big) (a_k+a_\ell) (b_m+b_n)/4 }{ \Tr ( E \rho ) } \notag \\
	& \qquad = \frac14 \frac{ \Tr ( E \hat{B} \hat{A} \rho ) + \Tr ( E \rho \hat{A} \hat{B} ) + \Tr ( E \hat{A} \rho \hat{B} ) + \Tr ( E \hat{B} \rho \hat{A} ) }{ \Tr ( E \rho ) } \notag \\
	& \qquad = \frac12 \left[ \re \left( \frac{\Tr (E\hat{B}\hat{A}\rho)}{\Tr (E\rho)} \right) + \re \left( \frac{\Tr (E\hat{A}\rho\hat{B})}{\Tr (E\rho)} \right) \right]
	 . \label{eq:x1x2}
\end{align}
For pure pre- and post-selected states (i.e.\ for $\rho = \ketbra{\psi}{\psi}$ and $E = \ketbra{\phi}{\phi}$), one has $\frac{\Tr (E\hat{A}\rho\hat{B})}{\Tr (E\rho)} = \frac{\braket{\phi|\hat{A}|\psi}\!\braket{\psi|\hat{B}|\phi}}{\braket{\phi|\psi}\!\braket{\psi|\phi}} = A_\psi^\phi (B_\psi^\phi)^*$, so that one recovers Eq.~\eqref{eq:SeqWeakMesOutcome}. Note however that Eq.~\eqref{eq:SeqWeakMesOutcome} does \emph{not} hold for the generalised weak values $(BA)_\rho^E$, $A_\rho^E$ and $B_\rho^E$; the correct generalisation is instead given by the equation above.

Similar calculations as above also lead to
\begin{align}
	\braket{\hat{p}_1 \otimes \hat{x}_2} & \ \approx \ \frac{1}{2\sigma_1^2} \, \frac12 \left[ \im \left( \frac{\Tr (E\hat{B}\hat{A}\rho)}{\Tr (E\rho)} \right) + \im \left( \frac{\Tr (E\hat{A}\rho\hat{B})}{\Tr (E\rho)} \right) \right], \label{eq:p1x2} \\
	\braket{\hat{x}_1 \otimes \hat{p}_2} & \ \approx \ \frac{1}{2\sigma_2^2} \, \frac12 \left[ \im \left( \frac{\Tr (E\hat{B}\hat{A}\rho)}{\Tr (E\rho)} \right) - \im \left( \frac{\Tr (E\hat{A}\rho\hat{B})}{\Tr (E\rho)} \right) \right], \label{eq:x1p2} \\
	\braket{\hat{p}_1 \otimes \hat{p}_2} & \ \approx \ -\frac{1}{4\sigma_1^2\sigma_2^2} \, \frac12 \left[ \re \left( \frac{\Tr (E\hat{B}\hat{A}\rho)}{\Tr (E\rho)} \right) - \re \left( \frac{\Tr (E\hat{A}\rho\hat{B})}{\Tr (E\rho)} \right) \right]. \label{eq:p1p2}
\end{align}

The real and imaginary parts of the weak value $(BA)_\rho^E = \frac{\Tr (E\hat{B}\hat{A}\rho)}{\Tr (E\rho)}$ are thus not directly given by the mean values of the pointer positions and momenta (conditioned on a successful post-selection), as observed previously in Refs.~\cite{Resch04,Mitchison07}, but can still easily be recovered by combining the mean values above as follows:
\begin{align}
\braket{\hat{x}_1 \otimes \hat{x}_2} - 4\sigma_1^2\sigma_2^2 \, \braket{\hat{p}_1 \otimes \hat{p}_2} \approx \re [(BA)_\rho^E], \qquad 2\sigma_1^2 \, \braket{\hat{p}_1 \otimes \hat{x}_2} + 2\sigma_2^2 \, \braket{\hat{x}_1 \otimes \hat{p}_2} \approx \im [(BA)_\rho^E]. \label{eq:recover_seqWV}
\end{align}

\bigskip

Nevertheless, if no post-selection is made ($E = \id$), then the two summands in Eq.~\eqref{eq:x1x2} and in Eq.~\eqref{eq:p1x2} are equal, and one again directly obtains
\begin{equation}
\braket{\hat{x}_1 \otimes \hat{x}_2} \approx \re [(BA)_\rho^{\idmini}] \, , \qquad \braket{\hat{p}_1 \otimes \hat{x}_2} \approx \frac{1}{2\sigma_1^2} \im [(BA)_\rho^{\idmini}] \label{eq:x1x2_p1x2_noPS}
\end{equation}
(as in Eq.~\eqref{eq:SeqWeakMesOutcomeNoPS} for $\braket{\hat{x}_1 \otimes \hat{x}_2}$).
The example of an imaginary anomalous weak value without post-selection using Pauli observables described in the main text, for which $(\hat\sigma_\textsc{x}\hat\sigma_\textsc{y})_{0}^{\idmini}=i$, can, from Eq.~\eqref{eq:x1x2_p1x2_noPS}, thus be observed by measuring $\braket{\hat{p}_1\otimes\hat{x}_2}$.

Moreover, these expressions hold as long as the weak regime is applicable for the first measurement, irrespective of the strength of the second~\cite{bednorz10,bednorz13}.
To see this, note that, in the absence of post-selection, Eq.~\eqref{eq:eta} reduces to
\begin{equation}
	\eta = \sum_{k\ell m}  \braket{b_m|a_k}\!\!\bra{a_k} \rho \ket{a_\ell }\!\!\braket{a_\ell |b_m} \ketbra{\varphi_1(a_k)}{\varphi_1(a_\ell )} \otimes \ketbra{\varphi_2(b_m)}{\varphi_2(b_m)}.
\end{equation}
Eq.~\eqref{eq:gaussian_approx} thus holds with \emph{exact equalities} for the terms in $\Tr(\hat{x}_1\otimes\hat{x}_2\, \eta)$ and $\Tr(\hat{p}_1\otimes\hat{x}_2\, \eta)$ corresponding to the second pointer, and the weak regime approximation is thus only required for the first measurement.
This also justifies further the claim that the second measurement can be seen as performing an effective post-selection, since it can be taken to be arbitrarily strong.

One can also calculate the variances of these quantities using the weak regime approximations given in Eqs.~\eqref{eq:gaussianx2} and~\eqref{eq:gaussianp2}, although the explicit expressions become somewhat complicated and are of insufficient interest to reproduce fully here.
We note simply that, in general, the variance of $\hat{x}_1\otimes\hat{x}_2$ is dominated by $\sigma_1^2\sigma_2^2$ in the weak regime so that more statistics are needed to determine the average product of pointer positions than for a single weak measurement.
However, in the absence of post-selection one has $(\Delta(\hat{x}_1\otimes\hat{x}_2))^2 \approx \sigma_1^2\sigma_2^2 + \sigma_1^2\braket{\hat{B}^2} + \sigma_2^2\braket{\hat{A}^2} + O[(\Lambda_\text{max}-\Lambda_\text{min})^2]$ (with $\Lambda_{\text{max/min}}=\Lambda_{\text{max/min}}(\hat{A},\hat{B})$ as defined in the main text), which again is valid as long as the weak regime is applicable for the first measurement.
Thus, by making the second measurement sufficiently strong (i.e., $\sigma_2$ small) the variance can be made to scale as $\sigma_1^2$, justifying our comment in Footnote~\ref{fn:scaling}.

\bigskip

Note that in some specific cases, the expectation values above can also be calculated exactly. For instance, in the first illustrative example (with no post-selection) introduced in the main text, we had
\begin{align}
\braket{\hat{x}_1 \otimes \hat{x}_2} &= \bra{\Phi_1^{(1)}} \hat{x}_1 \ket{\Phi_1^{(1)}} \bra{\varphi_2(1)} \hat{x}_2 \ket{\varphi_2(1)} + \bra{\Phi_1^{(0)}} \hat{x}_1 \ket{\Phi_1^{(0)}} \bra{\varphi_2(0)} \hat{x}_2 \ket{\varphi_2(0)} = \bra{\Phi_1^{(1)}} \hat{x}_1 \ket{\Phi_1^{(1)}}.
\end{align}
Using the explicit forms of Eq.~\eqref{Phi_10} for $\ket{\Phi_1^{(1)}}$ together with Eq.~\eqref{eq:gaussianpointer} for the first Gaussian pointer, one finds 
\begin{equation}
\braket{\hat{x}_1 \otimes \hat{x}_2} = \frac{1}{16} \big( 1 - 3 \braket{\varphi_1(1)|\hat{x}_1|\varphi_1(0)} - 3 \braket{\varphi_1(0)|\hat{x}_1|\varphi_1(1)} \big) = \frac{1}{16} \left( 1 - 3e^{-\frac{1}{8 \sigma_1^2}} \right),
\end{equation}
as in Eq.~\eqref{eq:x1x2_gaussian}. This is in indeed consistent with Eq.~\eqref{eq:x1x2_p1x2_noPS} in the weak regime limit where $\sigma_1 \gg 1$ (see Eq.~\eqref{eqn:ExampleWV}).

The explicit forms of Eqs.~\eqref{stateafterfirstmeasurement}, \eqref{jointpointerstate} and~\eqref{Phi_10} together with Eq.~\eqref{eq:gaussianpointer} also allow one to calculate the average positions of the two pointers independently. 
Specifically, one can calculate $\braket{\hat{x}_1}$ directly from the state after the first measurement, Eq.~\eqref{stateafterfirstmeasurement}, as the second measurement does not interact with the first pointer at all.
Observing that only the first term from Eq.~\eqref{stateafterfirstmeasurement} contributes (because $\bra{\varphi_1(0)}\hat{x}_1\ket{\varphi_1(0)} = 0$), we therefore have
\begin{align}
\braket{\hat{x}_1} &= \frac{1}{4} \braket{\psi_1|\psi_1} \braket{\varphi_1(1)|\hat{x}_1|\varphi_1(1)} \braket{\varphi_2(0)|\varphi_2(0)} = \frac14 \,. \notag \\
\end{align} 
For the second pointer, since $\bra{\varphi_2(0)}\hat{x}_2 \ket{\varphi_2(0)} = 0$, we find from Eqs.~\eqref{jointpointerstate}, \eqref{Phi_10} and~\eqref{eq:gaussianpointer} that
\begin{align}
\braket{\hat{x}_2} &= \braket{\Phi_1^{(1)}|\Phi_1^{(1)}} \braket{\varphi_2(1)|\hat{x}_2|\varphi_2(1)} = \braket{\Phi_1^{(1)}|\Phi_1^{(1)}} = \frac{1}{8} \left( 5 - 3e^{-\frac{1}{8\sigma_1^2}} \right) \in \left[ \frac{1}{4}, \frac{5}{8} \right] \,,
\end{align}
as claimed in the main text.

\subsection{Generalisation to $n$ sequential weak measurements}

Considering now $n$ sequential weak measurements, the previous definition of a weak value, for a system prepared in the (possibly mixed) state $\rho$ and post-selected on a POVM element $E$ can be generalised to 
\begin{equation}\label{eq:WVseq_general_n}
	(A_n \cdots A_1)_\rho^E \coloneqq \frac{\Tr (E\hat{A}_n \cdots \hat{A}_1\rho)}{\Tr (E\rho)}.
\end{equation}
For $\rho = \ketbra{\psi}{\psi}$ and $E = \ketbra{\phi}{\phi}$, this indeed reduces to the definition~\eqref{eq:weakvaluen} introduced in Ref.~\cite{Mitchison07}.
Without post-selection ($E = \id$), this simplifies to
$(A_n \cdots A_1)_\rho^{\idmini} \coloneqq \Tr (\hat{A}_n \cdots \hat{A}_1\rho)$, 
which in turn reduces to Eq.~\eqref{eq:weakvaluen_no_PS} for a pure pre-selected state $\rho = \ketbra{\psi}{\psi}$ -- and which, once again (and contrary to a general sequential weak value), is linear in the pre-selected state.%
\footnote{\parbox[t]{\textwidth}{Note that, as seen previously for one and two observables (cf.\ Footnotes~\ref{fn:reselection} and~\ref{fn:reselection2}), $(A_n \cdots A_1)_\psi^{\idmini} = (A_n \cdots A_1)_\psi^\psi$; that is, for a pure pre-selected state $\ket{\psi}$, performing no post-selection gives the same weak value as post-selecting on the pre-selected state, or ``re-selecting''~\cite{diosi16}. However, we emphasise that, as seen in Eqs.~\eqref{eq:SeqWeakMesOutcome} and~\eqref{eq:SeqWeakMesOutcomeNoPS}, the mean pointer positions will nonetheless differ in these physically distinct scenarios. Note also that this property does not generalise to mixed states.}}

\bigskip

The previous calculations for the mean values of products of positions and momenta can easily be generalised to $n$ sequential weak measurements.
For two complementary subsets ${\cal X}$ and ${\cal P}$ of $\{1, \ldots, n\}$, Eqs.~\eqref{eq:x1x2}--\eqref{eq:p1p2} generalise to (see also Ref.~\cite{Mitchison07})
\begin{align}
	\Big\langle \bigotimes_{i \in {\cal X}} \hat{x}_i \bigotimes_{j \in {\cal P}} \hat{p}_j \Big\rangle &\approx (-i)^{|{\cal P}|} \Big( \prod_{j \in {\cal P}} \frac{1}{2\sigma_j^2} \Big) \frac{1}{2^n} \frac{ \sum_{s_1, \ldots, s_n \in \{0,1\}} (-1)^{\sum_{j \in {\cal P}} s_j} \Tr ( E \hat{A}_n^{1-s_n} \cdots \hat{A}_1^{1-s_1} \rho \hat{A}_1^{s_1} \cdots \hat{A}_n^{s_n} ) }{ \Tr ( E \rho ) } .
\end{align}
After a couple of lines of algebraic manipulations (using in particular the fact that $\Tr ( E Q )^* = \Tr ( E Q^\dagger )$ for any matrix $Q$ of the form $\hat{A}_n^{1-s_n} \cdots \hat{A}_1^{1-s_1} \rho \hat{A}_1^{s_1} \cdots \hat{A}_n^{s_n}$, and relabelling certain summation indices $s_j \leftrightarrow 1-s_j$), this can be written as
\begin{align}
	\Big\langle \bigotimes_{i \in {\cal X}} \hat{x}_i \bigotimes_{j \in {\cal P}} \hat{p}_j \Big\rangle \approx & (-1)^{\lfloor \! \frac{|{\cal P}|}{2} \! \rfloor} \Big( \prod_{j \in {\cal P}} \frac{1}{2\sigma_j^2} \Big) \frac{1}{2^{n-1}} \notag\\
	& \times \sum_{s_2, \ldots, s_n \in \{0,1\}} (-1)^{\sum_{j \in {\cal P} \backslash \{1\}} s_j} \ {}^{\re\!}/_{\!\im} \Big[ \frac{\Tr ( E \hat{A}_n^{1-s_n} \cdots \hat{A}_2^{1-s_2} \hat{A}_1 \rho \hat{A}_2^{s_2} \cdots \hat{A}_n^{s_n} ) }{\Tr ( E \rho )} \Big], \label{eq:xX_pP}
\end{align}
with ${}^{\re\!}/_{\!\im}$ to be replaced by $\re$ if $|{\cal P}|$ is even, and by $\im$ if $|{\cal P}|$ is odd.

Generalising Eq.~\eqref{eq:recover_seqWV}, the real and imaginary parts of the sequential weak value $(A_n \cdots A_1)_\rho^E$ defined above are then recovered by combining the mean values $\big\langle \bigotimes_{i \in {\cal X}} \hat{x}_i \bigotimes_{j \in {\cal P}} \hat{p}_j \big\rangle$ as follows:
\begin{align}
	\sum_{{\cal P}: |{\cal P}| \, {}^{\text{even}\!}/{}_{\!\text{odd}}} (-1)^{\lfloor \! \frac{|{\cal P}|}{2} \! \rfloor} \big( \prod_{j \in {\cal P}} 2\sigma_j^2 \big) \ \Big\langle \bigotimes_{i \in {\cal X}} \hat{x}_i \bigotimes_{j \in {\cal P}} \hat{p}_j \Big\rangle 
	& \ \approx \ {}^{\re\!}/_{\!\im} \big[ (A_n \cdots A_1)_\rho^E \big], \label{eq:recover_seqWV_n}
\end{align}
where the sum is over the $2^{n-1}$ subsets ${\cal P}$ of $\{1,\ldots,n\}$ either such that $|{\cal P}|$ is even (for the real part), or such that $|{\cal P}|$ is odd (for the imaginary part), and with ${\cal X} = \{1,\ldots,n\} \backslash {\cal P}$.

\bigskip

In the case with no post-selection ($E = \id$) and when $n \notin {\cal P}$, Eq.~\eqref{eq:xX_pP} can be further simplified (using the cyclic property of the trace) to
\begin{align}
	\Big\langle \bigotimes_{i \in {\cal X}} \hat{x}_i \bigotimes_{j \in {\cal P}} \hat{p}_j \Big\rangle \approx & (-1)^{\lfloor \! \frac{|{\cal P}|}{2} \! \rfloor} \Big( \prod_{j \in {\cal P}} \frac{1}{2\sigma_j^2} \Big) \frac{1}{2^{n-2}} \notag\\
	&\times \sum_{\substack{s_2, \ldots, \qquad \\ \ s_{n-1} \in \{0,1\}}} \!\!\!\! (-1)^{\sum_{j \in {\cal P} \backslash \{1\}} s_j} \ {}^{\re\!}/_{\!\im} \big[ \Tr ( \hat{A}_2^{s_2} \cdots \hat{A}_{n-1}^{s_{n-1}} \hat{A}_n \hat{A}_{n-1}^{1-s_{n-1}} \cdots \hat{A}_2^{1-s_2} \hat{A}_1 \rho ) \big]. \label{eq:xX_pP_noPS}
\end{align}
Noting that $\Tr ( \hat{A}_2^{s_2} \cdots \hat{A}_{n-1}^{s_{n-1}} \hat{A}_n \hat{A}_{n-1}^{1-s_{n-1}} \cdots \hat{A}_2^{1-s_2} \hat{A}_1 \rho ) = ( A_2^{s_2} \cdots A_{n-1}^{s_{n-1}} A_n A_{n-1}^{1-s_{n-1}} \cdots A_2^{1-s_2} A_1)_\rho^{\idmini}$, one can see that $\big\langle \bigotimes_{i \in {\cal X}} \hat{x}_i \bigotimes_{j \in {\cal P}} \hat{p}_j \big\rangle$ is obtained in Eq.~\eqref{eq:xX_pP_noPS} as the real or imaginary part of a combination of $2^{n-2}$ sequential weak values with no post-selection, for $2^{n-2}$ different permutations of the observables.
The expression above allows one, in the case where $n=3$, ${\cal P} = \emptyset$ and $\rho = \ketbra{\psi}{\psi}$, to recover Eq.~\eqref{eq:Pointer3measurements} of the main text.
As for the case with two sequential weak measurements without post-selection, this relation between pointer expectation values and sequential weak values holds even if the final measurement is arbitrarily strong, and the weak regime need only be applicable for the first $n-1$ measurements~\cite{bednorz10,bednorz13}.

When $n \in {\cal P}$, $\big\langle \bigotimes_{i \in {\cal X}} \hat{x}_i \bigotimes_{j \in {\cal P}} \hat{p}_j \big\rangle$ vanishes to the first order in the absence of post-selection and the right-hand side of Eq.~\eqref{eq:xX_pP} becomes 0.
It follows that for $E = \id$, one can sum only over subsets ${\cal P}$ of $\{1,\ldots,n-1\}$ in Eq.~\eqref{eq:recover_seqWV_n} to recover the weak value $(A_n \cdots A_1)_\rho^E$.

Finally, when $\mathcal{P}=\emptyset$ and $E=\id$, $\braket{\hat{x}_1\otimes\cdots\otimes\hat{x}_n}$ can be expressed in a particularly nice form as the nested anti-commutator~\cite{bednorz10,diosi16}
\begin{equation}
	\braket{\hat{x}_1\otimes\cdots\otimes\hat{x}_n} \approx \frac{1}{2^{n-1}}\Tr[\{A_1,\{A_{2},\dots,\{A_{n-1},A_n\}\dots\}\}\,\rho\,].
\end{equation}

\section{Proof that $\re [(BA)_\psi^{\protect\idmini}] \ge -1/8$ for two projection observables $\hat{A}$ and $\hat{B}$ (with eigenvalues $0, 1$)}

Suppose the two observables that are sequentially weakly measured on some initial state $\ket{\psi}$ are projectors, $\hat{A}$ and $\hat{B}$.
Define then the (normalised) states $\ket{\alpha} \coloneqq \frac{\hat{A}\ket{\psi}}{\sqrt{\braket{\psi | \hat{A} | \psi}}}$ and $\ket{\beta} \coloneqq \frac{\hat{B}\ket{\psi}}{\sqrt{\braket{\psi | \hat{B} | \psi}}}$ (in the case where $\braket{\psi | \hat{A} | \psi} = 0$ or $\braket{\psi | \hat{B} | \psi} = 0$, one can define $\ket{\alpha}$ or $\ket{\beta}$ to be any state orthogonal to $\ket{\psi}$), such that $\braket{\alpha|\psi} = \sqrt{\braket{\psi | \hat{A} | \psi}} \ge 0$, $\braket{\beta|\psi} = \sqrt{\braket{\psi | \hat{B} | \psi}} \ge 0$, $\hat{A}\ket{\psi} = \braket{\alpha|\psi}\ket{\alpha}$ and $\hat{B}\ket{\psi} = \braket{\beta|\psi}\ket{\beta}$.
We then have:
\begin{align}
\re [(BA)_\psi^{\idmini}] = \re [\braket{\psi|\hat{B}\hat{A}|\psi}] = \braket{\alpha|\psi} \braket{\beta|\psi} \re [\braket{\beta|\alpha}]. \label{eq:proof18}
\end{align}

Now, using the AM-GM and Cauchy-Schwartz inequalities,
\begin{align}
\braket{\alpha|\psi} \braket{\beta|\psi} \le \left( \frac{\braket{\alpha|\psi} + \braket{\beta|\psi}}{2}\right)^2  = \left( \Big(\frac{\bra{\alpha} + \bra{\beta}}{2}\Big)\ket{\psi} \right)^2 \le \left\lVert\frac{\ket{\alpha} + \ket{\beta}}{2}\right\rVert^2 = \frac{1 + \re [\braket{\beta|\alpha}]}{2}.
\end{align}
Hence, either $\re [\braket{\beta|\alpha}] \ge 0$ in Eq.~\eqref{eq:proof18}, in which case (recalling that $\braket{\alpha|\psi} \braket{\beta|\psi} \ge 0$) one has $\re [(BA)_\psi^{\idmini}] \ge 0\, ( \ge -\frac{1}{8})$, or $\re [\braket{\beta|\alpha}] < 0$ and one then has (using the fact that $\frac{1+x}{2}x \ge -\frac{1}{8}$ for all $x$)
\begin{align}
\re [(BA)_\psi^{\idmini}] \ge \frac{1 + \re [\braket{\beta|\alpha}]}{2} \re [\braket{\beta|\alpha}] \ge -\frac{1}{8}\,,
\end{align}
which concludes the proof.

\medskip

Note, furthermore, that by the linearity of the weak value with no post-selection with respect to the pre-selected state, $(BA)_\rho^{\protect\idmini} = \sum_i q_i (BA)_{\psi_i}^{\protect\idmini}$ for a mixed state $\rho = \sum_i q_i \ketbra{\psi_i}{\psi_i}$, which implies that the bound above also holds for $(BA)_\rho^{\protect\idmini}$; that is, $\re [(BA)_\rho^{\protect\idmini}] \ge -1/8$ for any two projection observables $\hat{A}$ and $\hat{B}$ and any mixed state $\rho$.

\section{Proof of Eq.~\eqref{eqn:ampliBoundNobs}}

To bound the magnitude of the sequential weak value with no post-selection for $n$ measurements, let us write (using the Cauchy-Schwartz inequality in the first line):
\begin{align}
	|(A_n\cdots A_1)_\psi^{\idmini}|^2 \ = \ |\braket{\psi|\hat{A}_n\cdots \hat{A}_1|\psi}|^2 \ & \le \  \braket{\psi|\psi} \braket{\psi|(\hat{A}_n\cdots \hat{A}_1)^\dagger \hat{A}_n\cdots \hat{A}_1 |\psi} \notag\\[1mm]
		& = \, \braket{\psi|\hat{A}_1\cdots \hat{A}_{n-1}\hat{A}_n^2\hat{A}_{n-1}\cdots \hat{A}_1|\psi} \notag \\[-1mm]
		& \le \, \lVert\hat{A}_n\rVert^2\braket{\psi|\hat{A}_1\cdots \hat{A}_{n-1}^2\cdots \hat{A}_1|\psi} \ \le \ \ \cdots \ \ \le \ \prod_{j=1}^n \lVert\hat{A}_j\rVert^2.
\end{align}

\medskip

We further note that by using the linearity of the weak value with no post-selection and the triangle inequality, it is easy to see that Eq.~\eqref{eqn:ampliBoundNobs} also holds for a preparation in any mixed state $\rho = \sum_i q_i \ketbra{\psi_i}{\psi_i}$:
\begin{align}
	|(A_n\cdots A_1)_\rho^{\idmini}| \ = \ \big| \sum_i q_i (A_n\cdots A_1)_{\psi_i}^{\idmini} \big| \ & \le \ \sum_i q_i |(A_n\cdots A_1)_{\psi_i}^{\idmini}| \ \le \ \big( \underbrace{{\textstyle \sum_i q_i}}_{1}\big) \, \prod_{j=1}^n \lVert\hat{A}_j\rVert \, .
\end{align}

\end{document}